\documentclass[twocolumn,]{aastex61}
\usepackage{amssymb,amsmath}
\usepackage[T1]{fontenc}
\usepackage[utf8x]{inputenc}
\usepackage{listings}
\usepackage{graphicx}
\hypersetup{breaklinks=true,
           bookmarks=true,
           pdfauthor={},
           pdftitle={Neutral ISM, Lyman-Alpha and Lyman-continuum in the nearby starburst Haro 11},
           colorlinks=true,
           citecolor=blue,
           urlcolor=blue,
           linkcolor=magenta,
           pdfborder={0 0 0}}
\shorttitle{Neutral ISM, Ly-Alpha and LyC in Haro 11}

\shortauthors{T. E. Rivera-Thorsen et al.}

\usepackage[caption=false]{subfig}

\submitjournal{\apj}
\received{November 24, 2016}
\revised{January 25, 2017}
\accepted{January 26, 2017}
\begin{document}

\title{Neutral ISM, Lyman-Alpha and Lyman-continuum in the nearby starburst Haro
11 \footnote{Based on observations with HST-COS, program GO 13017, obtained from
the Mikulski Archive for Space Telescopes (MAST). STScI is operated by the
Association of Universities for Research in Astronomy, Inc., under NASA contract
NAS5-26555. Support for MAST for non-HST data is provided by the NASA Office of
Space Science via grant NNX09AF08G and by other grants and contracts.} }

\correspondingauthor{T. Emil Rivera-Thorsen}
\email{trive@astro.su.se}

\author{T. Emil Rivera-Thorsen}
\affil{Department of Astronomy, Stockholm University, AlbaNova University
Centre, SE-106 91 Stockholm, Sweden}
\affil{Oscar Klein Centre for Cosmoparticle Physics, Stockholm, Sweden}
\author{Göran Östlin}
\affil{Department of Astronomy, Stockholm University, AlbaNova University
Centre, SE-106 91 Stockholm, Sweden}
\affil{Oscar Klein Centre for Cosmoparticle Physics, Stockholm, Sweden}
\author{Matthew Hayes}
\affil{Department of Astronomy, Stockholm University, AlbaNova University
Centre, SE-106 91 Stockholm, Sweden}
\affil{Oscar Klein Centre for Cosmoparticle Physics, Stockholm, Sweden}
\author{Johannes Puschnig}
\affil{Department of Astronomy, Stockholm University, AlbaNova University
Centre, SE-106 91 Stockholm, Sweden}
\affil{Oscar Klein Centre for Cosmoparticle Physics, Stockholm, Sweden}

\begin{abstract} 
		
Star forming galaxies are believed to be a major source of Lyman Continuum (LyC)
radiation responsible for reionizing the early Universe. Direct observations of
escaping ionizing radiation have however been few and with low escape fractions.
In the local Universe, only ~10 emitters have been observed, with typical escape
fractions of a few percent. The mechanisms regulating this escape need to be
strongly evolving with redshift in order to account for the Epoch of
Reionization. Gas content and star formation feedback are among the main
suspects, known to both regulate neutral gas coverage and evolve with cosmic
time. In this paper, we reanalyze HST-COS spectrocopy of the first detected
local LyC leaker, Haro 11. We examine the connection between LyC leakage and
Lyman-$\alpha$ line shape, and feedback-influenced neutral ISM properties like
kinematics and gas distribution. We discuss the two extremes of an optically
thin, density bounded ISM and a riddled, optically thick, ionization bounded
ISM, and how Haro 11 fits into their theoretical predictions.  We find that the
most likely ISM model is a clumpy neutral medium embedded in a highly ionized
medium with a combined covering fraction of unity and a residual neutral gas
column density in the ionized medium high enough to be optically thick to
Lyman-$\alpha$, but low enough to be at least partly transparent to Lyman
continuum and undetected in Si II. This suggests that SF feedback and
galaxy-scale interaction events play a major role in opening passageways for
ionizing radiation through the neutral medium. 

\end{abstract}

\section{Introduction and
Observations}\label{introduction-and-observations}

Young, star-forming galaxies are believed to be the source of a major part of
the radiation which reionized the early Universe. It is however unclear which
physical conditions can facilitate the escape of the necessary amount of
radiation, given that these galaxies contain large amounts of neutral gas which
is opaque to this ionizing radiation at column densities above $\log N \sim
17.2$ \citep{Verhamme2015}.  Searches for leaking galaxies at redshifts $z
\gtrsim 1$ have yielded few detections \citep[e.g.][]{Iwata2009, Vanzella2010,
Vanzella2012, Nestor2013, Cowie2009, Siana2010}, with escape fractions
well below the $\sim 20\%$ needed to account for cosmic reionization
\citep{Bouwens2011, Robertson2013}. A population of lower mass and lower
luminosity, star forming galaxies are expected to contribute to reionization,
but high star formation is usually coincident with higher neutral gas
(surface) density, which would cause a higher probability of blocking
the ionizing photons \citep[e.g.][]{Erb2016, Robertson2013}. In the local
Universe, only 9 leakers have been detected so far \citep{Bergvall2006,
Leitet2011, Leitet2013, Borthakur2014, Izotov2016Nat, Izotov2016MNRAS,
Leitherer2016}, with escape fractions ranging typically between 1-8\%,
with one as high as $f_{\mathrm{esc}} \approx 13\%$ \citep{Izotov2016MNRAS}.

Models of ISM surrounding a central source and allowing escape of Lyman
continuum (LyC), span the range of two extremes: In one regime, the surrounding
gas is optically thin, highly ionized and density-bounded \citep[see
e.g.][]{Jaskot2013}, allowing escape for at least a fraction of
ionizing photons. In the other regime, the central source is surrounded by an
optically thick, ionization bounded medium with the neutral medium surrounding
the central Strömgren sphere not completely covering all lines of sight to the
background source in what is called the \emph{picket fence model}
\citep{Conselice2000, Bergvall2006, Heckman2011, Zackrisson2013} or the
\emph{riddled ionization bounded medium} by \cite{Verhamme2015}.

The latter paper presents modeling of
the imprints of these two extreme scenarios on the observed spectral
signature of Lyman-$\alpha$ emission lines, and suggests
how these can help point to candidate LyC leakers. The authors compare their
theoretical predictions to sample of Lyman-$\alpha$ profiles, including
a section of the spectrum treated in this work.

\begin{figure}
\centering
\includegraphics[width=3.500in]{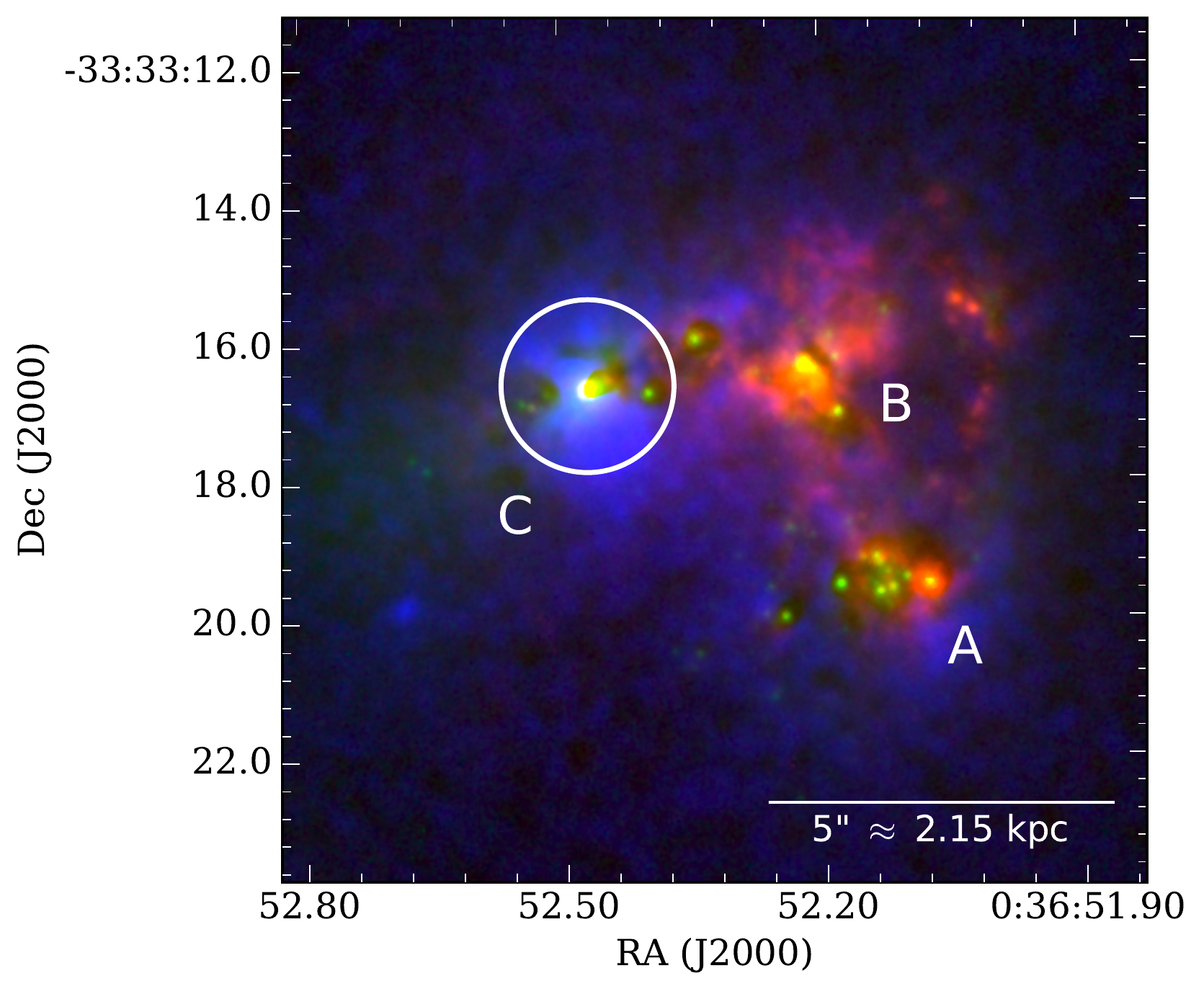}
\caption{Approximate position of the COS aperture, shown on HST imaging
data of \citet{Hayes2009}; \citet{Ostlin2009}, encoding UV continuum in
green, H$\alpha$ in red, and continuum subtracted Lyman $\alpha$ in
blue. N is up, E is to the left.}\label{fig:apert}
\end{figure}

The galaxy treated in this study is Haro 11, a well-studied Blue Compact Galaxy
at low redshift ($z = 0.021$). Morphologically, it is dominated by three major
star-forming knots, called knot A, B and C respectively, following the
terminology of \citet{Vader1993} \citep[see also][ and
fig.~\ref{fig:apert}]{Kunth1998, Hayes2007}.  Knots B and C are both very strong
in H$\alpha$, revealing that they also are producing strong Ly$\alpha$ 
But while knot C is a strong
Ly$\alpha$ emitter, knot B is a strong absorber, indicating that the neutral ISM
properties along the line of sight to the two knots are quite different
\citep{Hayes2009,Ostlin2009}.  Its complex kinematics indicates that it is
undergoing a merger or major interaction event \citep{Ostlin2001, Ostlin2015,
James2014}. It is the first galaxy in the local Universe reported to leak LyC
\citep{Bergvall2006}. The same dataset has since been re-analyzed by
\cite{Grimes2007} who concluded that there was no convincing evidence for LyC
escape; and since by \cite{Leitet2011} who report an escape fraction of $3.3 \pm
0.7 \%$ based on a new background modeling. In this work, we assume that some
radiation does escape, but it should be kept in mind that this result has been
disputed. Since Ly$\alpha$ escape is favored 
by some of the same conditions which also allow LyC escape 
\citep[see e.g. ][]{Hayes2015, DijkstraRev, Leitet2011}, the leaking photons 
are assumed to originate from knot C, an assumption adopted in this work.

In Rivera-Thorsen et al., submitted to \apj, we analyzed optical and NUV nebular
emission lines in slit spectra from ESO VLT/X-Shooter \citep[see
also][]{Guseva2012}. We find from kinematics modelling that both knot B and knot
C are associated to a component blueshifted by $\sim 50 - 100$ km s$^{-1}$
relative to the mean nebular velocity.  This component extends as far as $\sim
200$ pc. SE of knot B, and $\sim 1.5$ kpc S of the midpoint between the two
knots. Given the dense cloud coverage of knot B and the strong star formation
activity in both knots, we conclude that this component is not only approaching
but also found in front of the starburst regions and thus along the LOS to knot
C.

Recently, the galaxy has been observed in 21 cm \ion{H}{1} emission with
the 100 m. Robert C. Byrd Green Bank Telescope \citep{Pardy2016arXiv}.
The authors find that Haro 11 has a low gas mass and a very low neutral
gas to stellar mass fraction, and that it contains around twice as much
ionized as neutral Hydrogen. Interestingly, the authors also find that
the bulk of the neutral gas is redshifted relative to the systemic 
velocity defined from nebular emission from the \ion{H}{2} regions surrounding
the main starbursts, signifying that the majority of the neutral gas
reserves are detached from the star formation activity, possibly a tidal
arm being flung outwards as a part of the ongoing merger event.

In this paper, we re-analyze the HST-COS spectrum of Haro 11 C acquired
as part of HST program GO 13017, PI Timothy Heckman. It was first
published in \citet{Alexandroff2015} and \citet{Heckman2015} as part of
a sample of 22 Lyman Break Analog (LBA) galaxies analyzed individually
and as a stack. These authors mainly focus on three indirect indicators
of LyC leakage \citep{Overzier2009, Heckman2011}: Residual flux in the
trough of saturated ISM absorption lines, blueshifted emission in
Lyman-$\alpha$, and weak optical {[}\ion{S}{2}{]} emission lines.
Details about observation and data reduction are described in depth in
\citet{Alexandroff2015}; we point the reader there for further
information about these. The Ly$\alpha$ profile in this spectrum is, in
addition to \citet{Heckman2011}, also discussed in \citet{Verhamme2015}.

We measure a number of kinematic properties for both the neutral (Low-Ionized
State, LIS) and ionized (High-Ionized State, HIS) phase. We apply the apparent
optical depth method \citep[AOD, ][]{Savage1991, Pettini2002, Quider2009,
Jones2013}, with the implementation being as described in
\citet{RiveraThorsen2015} (RT15) to infer geometric properties of the neutral
medium. Under the assumtion that the LyC photons do indeed escape from knot C,
we constrain the column density of neutral hydrogen covering the background
source.

\begin{figure*}
\centering
\includegraphics[]{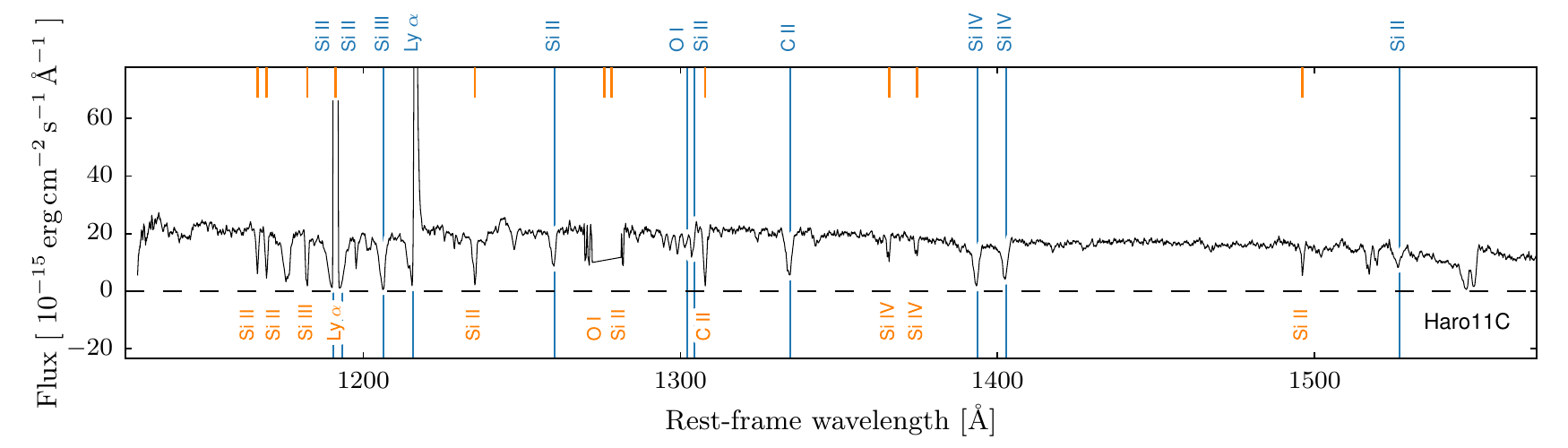}
\caption{Overview of the relevant regions of the COS spectrum of Haro 11
C, shown in rest frame wavelengths of the object. Some key spectral
lines are marked; Milky Way features in orange, features of Haro 11 in
blue.}\label{fig:fullspec}
\end{figure*}

Fig.~\ref{fig:fullspec} gives an overview of the COS spectrum of Haro 11
C treated in this work. Some important features are marked in blue
(internal to Haro 11) or orange (Milky Way features).

\subsection{Effective resolution}\label{effective-resolution}

For a point source, the resolution of the Cosmic Origins Spectrograph is
$R=20,000$, which corresponds to six pixels of the extracted spectrum
per resolution element. We have therefore binned the data by a factor of
6 to minimize noise while not losing information. The resolution,
however, is generally lower than this for extended sources; for a
uniformly filled aperture, it is as low as $R\approx2,000$. For more
morphologically complex sources, the effective resolution depends on the
exact shape and angular size of the target.

To estimate the effective resolution for this target, we proceeded as
follows. We used imaging data of Haro 11 in the UV-continuum
from \citet{Ostlin2009, Hayes2009}, since the resolution of absorption
lines is determined by the resolution of the background continuum source. We 
then extracted a circular image at the same
position and radius as the COS aperture, which was rotated to match the
orientation of the COS as given in the spectrum headers. We modelled
vignetting in the COS aperture by multiplying each pixel in the circular
UV-continuum image with an interpolation of the values given in the
throughput grid in Fig. 7 in \citet{CosImaging}. Finally, the aperture
cutouts were collapsed along the cross-dispersion direction, and the
FWHM of the flux distribution along the dispersion direction is reported
as the effective resolution of this observation. This is converted from
arc seconds to COS resolution elements by the factor 0.171 arcsec/resel
reported in the COS instrument handbook \citep{CosHandbook}

\begin{figure}
\centering
\includegraphics[width=1.000\hsize]{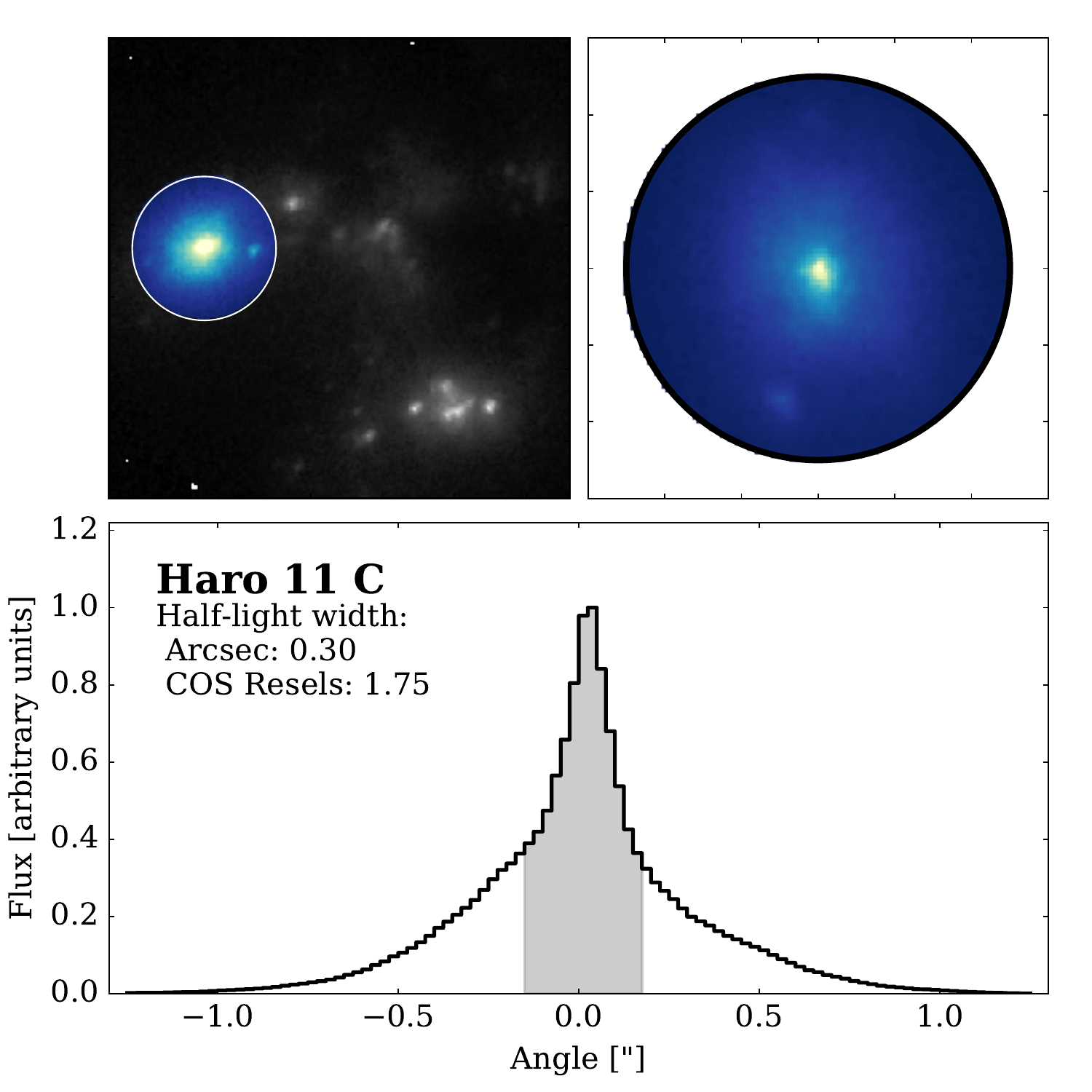}
\caption{Effective resolution estimate. \textbf{Upper left}: Haro 11 in
UV-continuum, data from \citet{Ostlin2009, Hayes2009}. Data is
shown on a square root scale and cut levels set for best detail. Inset
in blue is the COS aperture covering the region surrounding knot
B in the terminology of \citet{Vader1993}. N is up, E is left.
\textbf{Upper right}: The extracted, throughput-corrected in-aperture
image, rotated so horizontal corresponds to the dispersion direction of
the spectrograph. \textbf{Lower panel}: The collapsed flux profile in
the aperture, with the shaded region showing the width of the effective
resolution element, which is also given in arc seconds and resolution
elements/bins.}\label{fig:resol}
\end{figure}

The procedure is illustrated in Fig.~\ref{fig:resol}. Here, the upper left panel
shows Haro 11 in UV-continuum using HST imaging data from \citet{Ostlin2009};
\citet{Hayes2009}, with the COS aperture coverage from this observation overlaid
in blue. The upper right panel shows the galaxy as the COS saw it, with the
region covered by the aperture cut out, rotated and vignetting-corrected as
described above.  Finally, the lower panel shows the flux profile in the
aperture, collapsed along the cross dispersion direction. The FWHM of this
distribution has been adopted as the effective resolution for this observation.
We find $R_{\text{eff}} = 0.3 \arcsec = 1.75$ spectral resolution elements as
defined in the COS instrument handbook \cite{CosHandbook}, which at the
wavelength of observed Ly$\alpha$ corresponds to 31 km s⁻¹. This value might
however be slightly underestimated, since the flux profile is not well described
by a Gaussian profile, but has stronger wings due to the morphology of the
target in the aperture, meaning that a larger fraction of the total flux is
outside the FWHM.

\section{Analysis}\label{analysis}

\subsection{Individual lines}\label{individual-lines}

\begin{figure}
\centering
\includegraphics[width=3.500in]{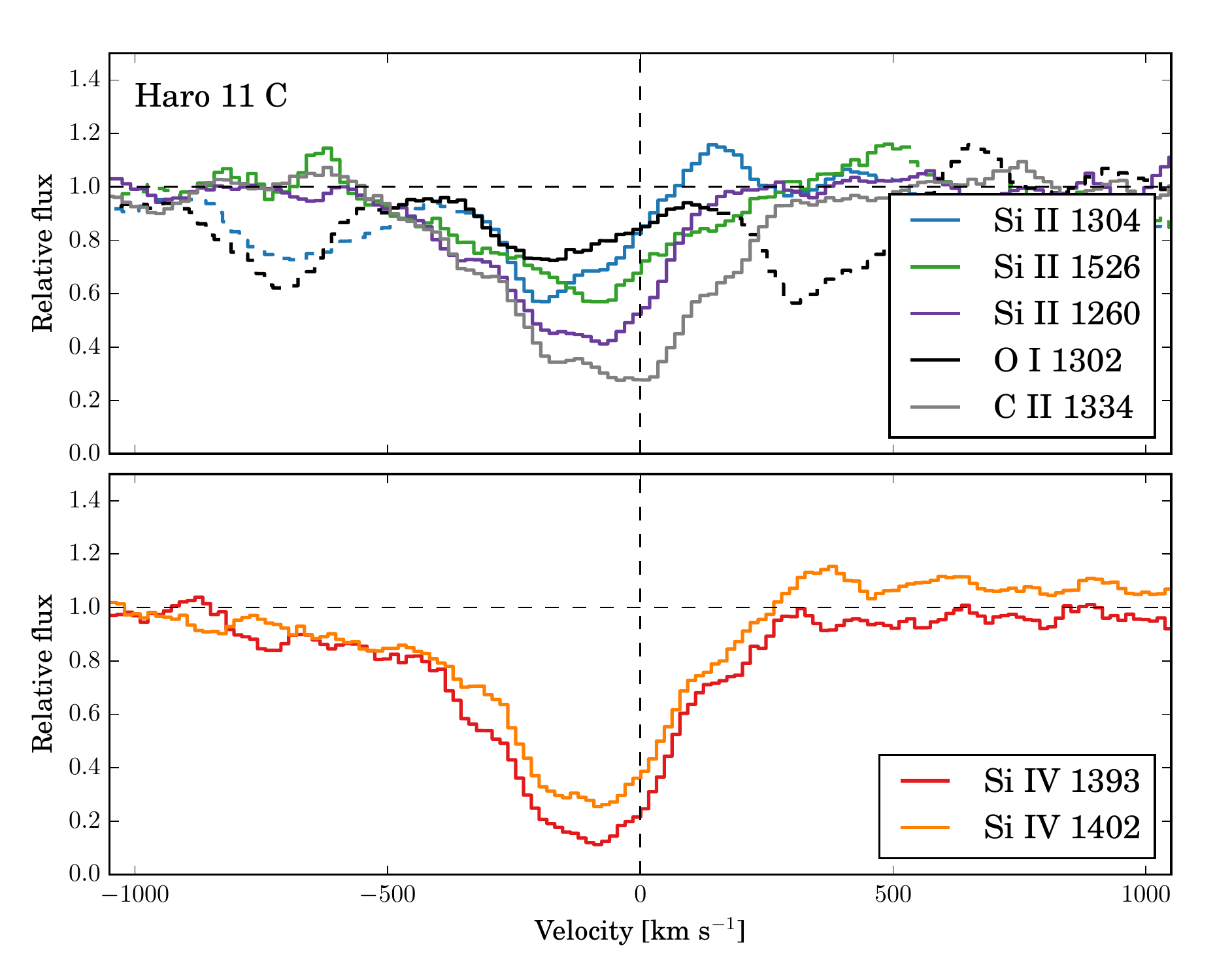}
\caption{The \ion{Si}{2} and other LIS (\textbf{upper}) and \ion{Si}{4}
(\textbf{lower}) profiles included in this study. Dashes denote data
points masked out due to foreground contamination. Data in this figure has
been smoothed to better show the line shapes, but all computations have been
carried out on the unsmoothed data. Note the
high-velocity component in both \ion{Si}{4} lines, extending out to
$\sim 850$~km~s$^{-1}$.}\label{fig:SingleLines}
\end{figure}

Figure~\ref{fig:SingleLines} shows the individual profiles of the
transitions included in our analysis; the upper panel shows transitions
of \ion{Si}{2}, lower panel of \ion{Si}{4}. It is plainly visible that
the ionization fraction is high, with the \ion{Si}{4} curves being
considerably deeper than the low-ionized lines. Looking at the upper
panel, \ion{Si}{2} $\lambda \lambda 1304, 1526$ are somewhat shallower
than \ion{Si}{2} $\lambda 1260$. The former two lines have comparable
oscillator strengths, both about a factor of 10 lower than that of
$\lambda 1260$. It is thus clear that we do not find ourselves in the
optically thin regime, in which the latter line should be
correspondingly around 10 times stronger; on the other hand, it is
possible that the two weak lines are not completely saturated. In the
lower panel, the two \ion{Si}{4} lines have oscillator strengths within
a factor of 2 of each other. They are thus at first glance consistent
with a medium that is not completely opaque, but not with an optically
thin one, and within uncertainties consistent with the optically thick.
The stronger absorption in \ion{Si}{4} reveals a high level of
ionization of the medium covering the background source.

\begin{deluxetable}{lC} 
  \tablecaption{Measured properties}
  \tablehead{\colhead{Quantity} & \colhead{Haro 11 C} }
  \startdata
    $z$ \tablenotemark{a}             & 0.02043 \pm 0.00002  \\
    $z_v$ [km s⁻¹]  \tablenotemark{a} & 6126 \pm 7           \\
    $\Delta v^{\rm LIS}$              & 480 \pm 15           \\
    $v_{\mathrm{int}}^{\rm LIS}$      & -149 \pm 7           \\
    $v_{95\%}^{\rm LIS}$              & -421 \pm 12          \\
     $v_{\rm peak}^{\rm Ly\alpha}$    & 158 \pm 0.8         \\
    $\log_{10}(N_{\mathrm{Si II}})^{v = 0}$  &  12.1 \pm  0.2 \\
  \enddata\
  \tablenotetext{\textrm a}{From \citet{Sandberg2013}}
\end{deluxetable}

\subsection{$N_{\rm Si}$ and $f_C$}\label{sec:aod}

Following the method described in RT15, we have performed fits for
column density and covering factor in each velocity bin, for both the
high- and low-ionization state. Here we shall briefly summarize the
method, but refer to RT15 and references therein for a detailed
explanation.

In any given velocity bin, the residual line intensity in terms of the
continuum intensity is given as
\begin{equation}
\label{eq:II0}
\frac{I}{I_0} = 1 - f_C (1 - e^{-\tau}),
\end{equation}
 with the optical depth $\tau$ given as:
\begin{equation}
\label{eq:tau}
\tau = f\lambda \frac{\pi e^2}{m_e c} N 
       = f\lambda \frac{N}{3.768 \times 10^{14}}
\end{equation}
 Here, $f$ is the oscillator strength of a given transition, $\lambda$
is its rest frame wavelength in Å, $N = N(v)$ is the column density of
the relevant ion \emph{within the given velocity bin}, and $f_{C}$ is
the covering fraction of neutral gas in same velocity bin. When multiple
absorption lines are present which arise from the same ground state; the
population of this state is the same for all transitions, and their
relative strengths are governed simply by their oscillator strengths,
and with two or more such transitions, $f_{C}$ and $N$ can be inferred
from knowledge of $f \lambda$ and measured values of $I/I_0$.

\begin{figure}
\centering
\includegraphics[width=0.800\hsize]{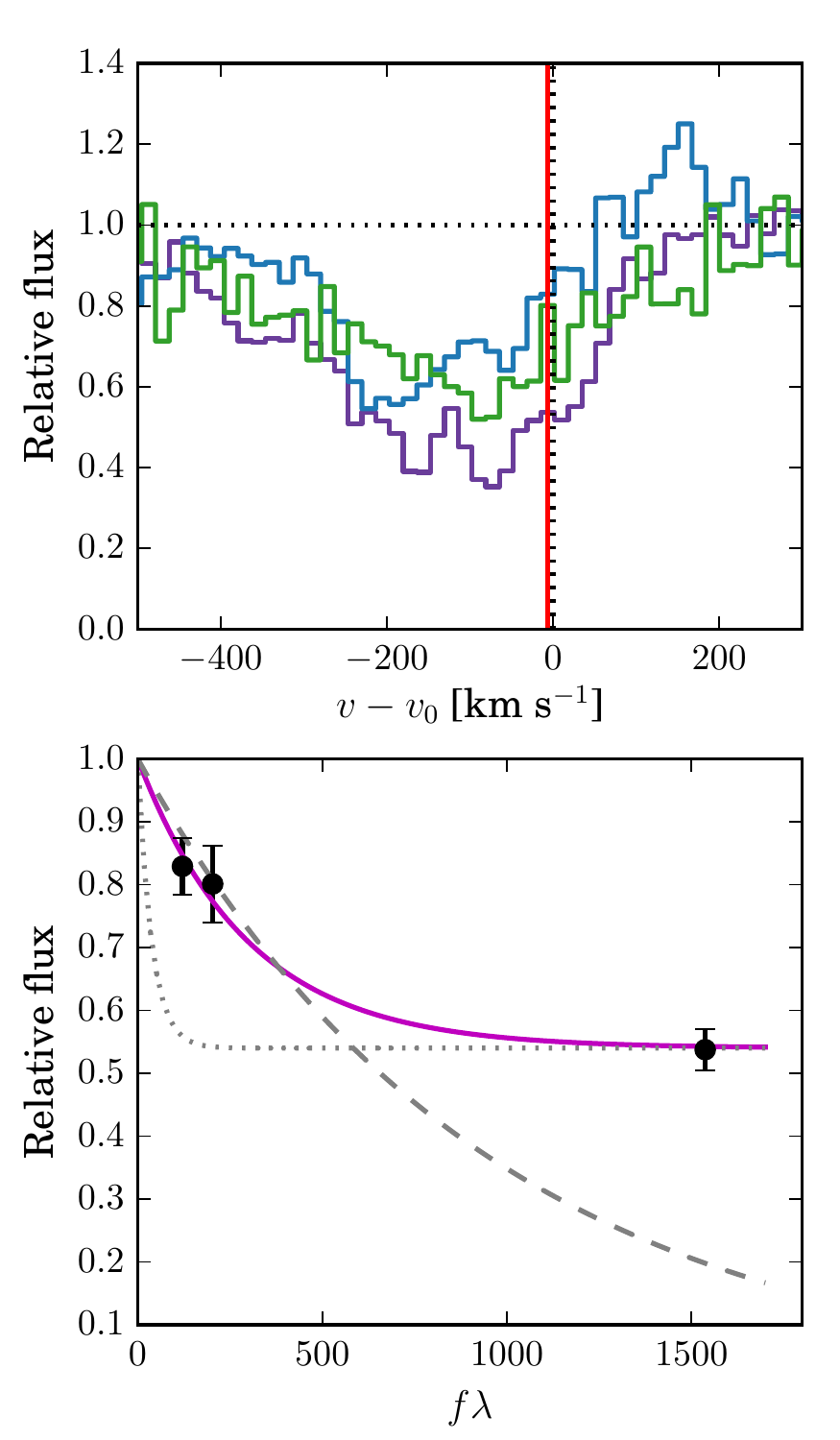}
\caption{Illustration of the Apparent Optical Depth computations. 
\textbf{Upper panel}: the three \ion{Si}{2} lines in this analysis shown 
together, with the zero-velocity bin marked with a vertical red line. 
\textbf{Lower panel}: From this bin, the three measured relative intensities are 
plotted against $f\lambda$ as black dots, with vertical error bars showing the
error spectrum from the COS pipeline. The best-fit
function $I/I_0 (f \lambda)$ as defined above is drawn in magenta. Also shown 
for illustration are the two extreme examples of a very high column density
and unchanged covering fraction, and of a very high covering fraction
but significantly lower column density.}\label{fig:AOD}
\end{figure}

The method is illustrated in Fig.~\ref{fig:AOD}: In the upper panel is
shown the line profiles of the three \ion{Si}{2} transitions included in
the analysis. The red vertical line marks the zero-velocity bin. From
this bin, the three relative fluxes are plotted in the lower frame
against their wavelength scaled oscillator strengths $f\lambda$ on the
$x$ axis. In magenta is shown the best fit of the function
$I/I_0 (f\lambda)$ described above. Also shown are two examples of
different parametrizations of this function, to show how the two
parameters influence its shape. One, in dotted gray, shows what happens
to the best-fit function when $N$ is raised strongly, while $f_C$ is
kept unchanged, while the other in gray dashes shows the curve for a
$f_C \approx 1$ and $N$ is $~0.5$ dex below the best-fit value. This is
of interest to the discussion of radiative transfer effects below in
sect.~\ref{sec:rt}.

\begin{figure*}
\centering
\subfloat[LIS\label{coldenLIS}]{\includegraphics[width=0.400\hsize]{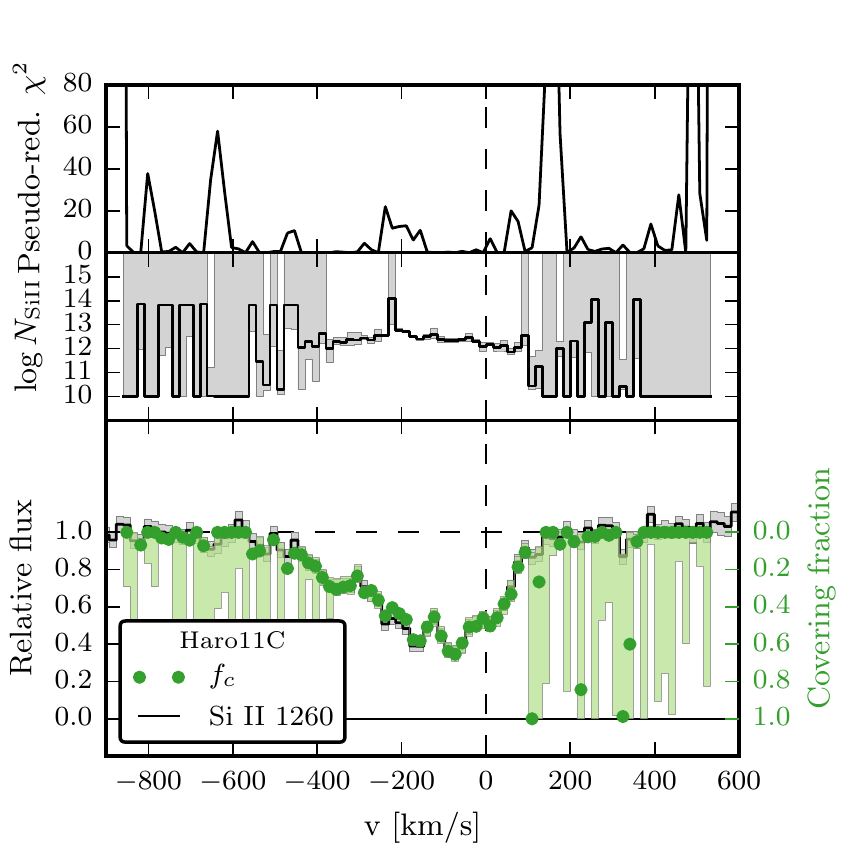}}
\subfloat[\ion{Si}{4}\label{coldenHIS}]{\includegraphics[width=0.400\hsize]{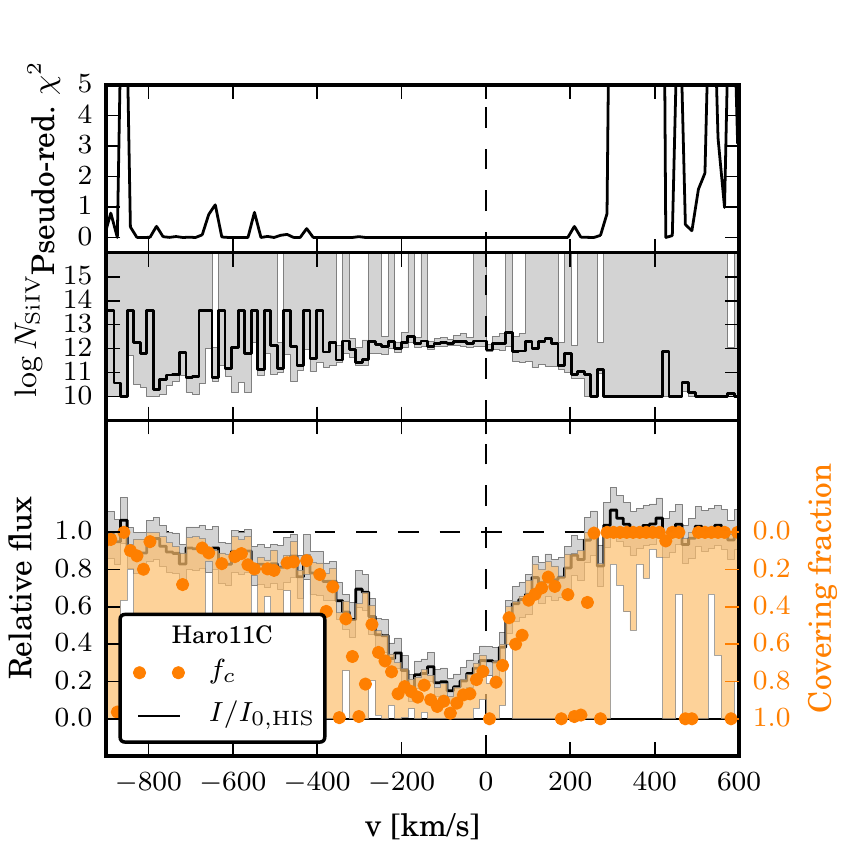}}
\caption{\textbf{Upper panels}: Pseudo-reduced $\chi^2$ as described in
RT15. \textbf{Middle panels}: Best-fit ion column density with
confidence intervals in shaded gray. \textbf{Lower panels}: \ion{Si}{2}
1260 / mean\ion{Si}{4} profile shown as black steps, with inferred $f_C$
shown with yellow dots. Lighter shaded columns show confidence intervals
for both.}\label{fig:WithColDens}
\end{figure*}

The resulting values of $N_{\text{Si\textsc{ii}}}$ and $f_C$ are shown
in fig.~\ref{fig:WithColDens}. The upper panels show the pseudo-reduced
$\chi^2$ as defined in RT15 ( $=\chi^2 / (\mathrm{DOF} + 1)$) for each
bin, middle panels show the inferred column density in each bin, with
surrounding shaded columns showing the confidence intervals. In the
lower panels, the mean LIS line profile is shown in black with gray
shaded uncertainty intervals. On these are overlaid the best-fit values
of $f_C$ as colored dots, with surrounding shaded bars showing the
confidence intervals. We again caution that $f_C$ is the covering
fraction of neutral gas \emph{within the given velocity bin}, and hence
only provides a lower limit for the total, geometric neutral gas
covering fraction, since gas at different velocities generally does not
occupy the same projected area.

\section{Discussion and conclusions}\label{discussion-and-conclusions}

\subsection{Lyman-$\alpha$ and ISM absorption
profiles}\label{sec:LISLya}

\begin{figure}
\centering
\includegraphics[width=3.500in]{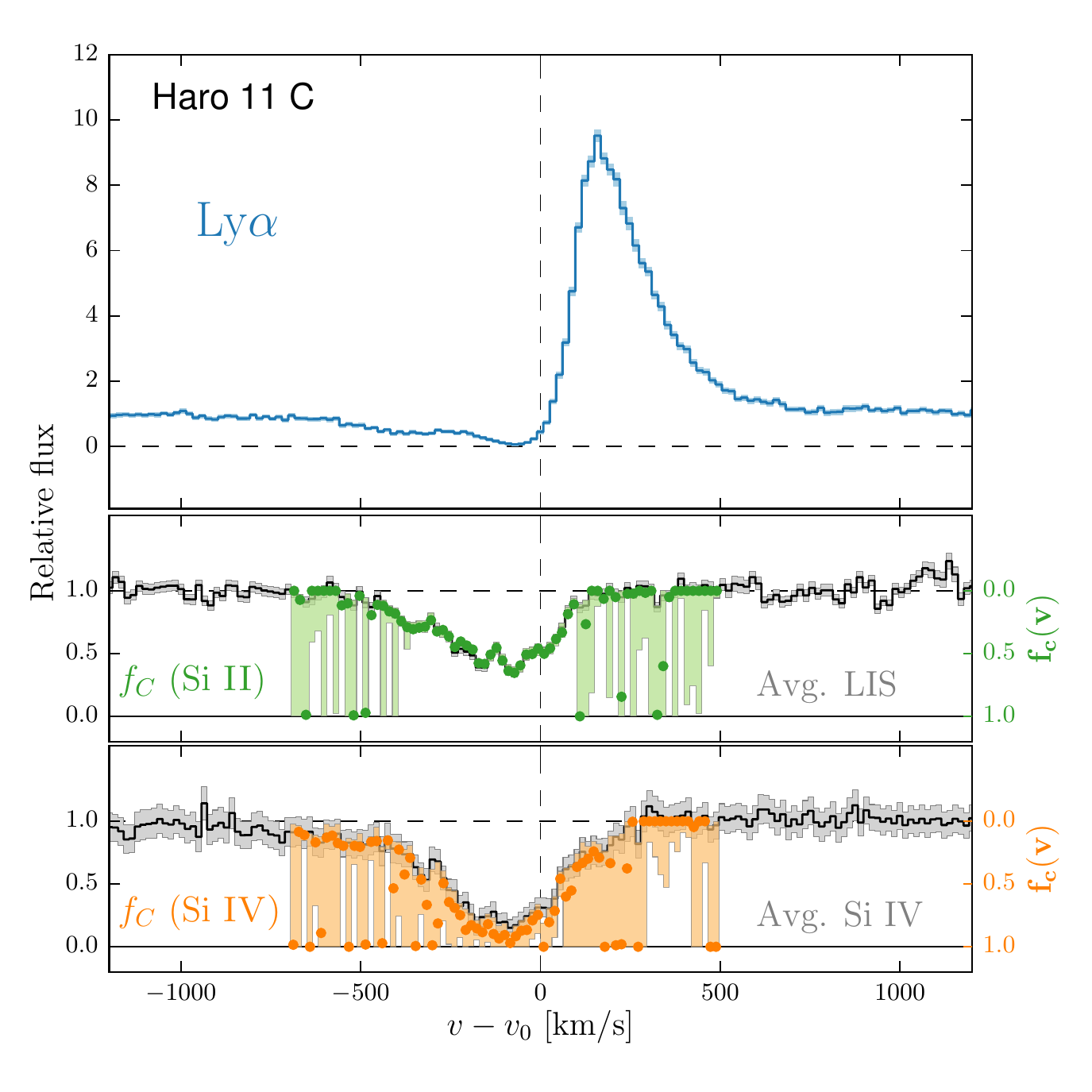}
\caption{\textbf{Upper panel}: Ly$\alpha$ profile of Haro 11 C, in
approximate units of the surrounding continuum level. The full line
indicates the measured values smoothed by a 5 px. flat kernel; surrounding shading
encloses the $\pm 1 \sigma$ error band (mostly comparable in width to
the line width). \textbf{Middle panel}: Black steps show the averaged
LIS line profile, smoothed by a 5px kernel. Surrounding gray shading
denotes the $\pm 1 \sigma$ confidence band. \textbf{Lower panel}: Same
as middle panel, but for the \ion{Si}{4} transitions.
}\label{fig:HisLisLya}
\end{figure}

In fig.~\ref{fig:HisLisLya}, we show the neutral and ionized absorption
profile as in fig.~\ref{fig:WithColDens} together with the profile of
Ly$\alpha$ on a common velocity scale.

The picture is what we would expect from a LyC leaker: The neutral absorption
features are weak, and the ionized features are strong, revealing a   highly
ionized medium in front of the central cluster . This is also in good agreement
with the findings of \citet{Pardy2016arXiv}, who conclude that the galaxy has
about twice as much ionized as neutral gas. In addition to this, it is
interesting to note the close similarity in shape between the \ion{Si}{2} and
\ion{Si}{4} line profiles, indicating that they likely represent two different
phases in the same higher density regions. These regions will likely be ionized
on the side facing the central cluster, being photoionized by its hot, massive
stars. This also implies that the nebular emission outflow found in
Rivera-Thorsen et al, submitted, at least partially traces the same gas as the
\ion{Si}{4} absorption here and, by extension, also the neutral medium. Looking
at fig.~\ref{fig:SingleLines}, we see that maximum absorption and thus the
largest amount of gas in a single velocity bin for both phases is found at
around $v \approx -50$ km s⁻¹. Interestingly, a component was also found in
nebular emission at this velocity in the observations of Rivera-Thorsen et al.,
submitted. The outflow velocity $v_{\text{int}}$ is at 149 ± 7 km s⁻¹, fully
consistent with the findings of \citet{Heckman2015} and \citet{Alexandroff2015}.
Interestingly, \citet{Sandberg2013} found from the neutral Sodium resonance
absorption doublet $\lambda \lambda 5889.95,5895.92$ (NaD) an overall, weak
\emph{redshift} of $v = 32$ km s⁻¹. From e.g.~fig.~\ref{fig:HisLisLya}, it is
evident that while neutral gas \emph{is} present at these velocities, the
velocity is at odds with the integrated velocity of $v = -149$ km s⁻¹ found in
this work. We note that \ion{Na}{1} has a very low ionization potential of only
$\sim 5$ eV, meaning that these atoms may well only be present in the densest
and/or dustiest regions, in which Sodium is shielded from ionization.  This is
in agreement with the finding of \citet{Sandberg2013} that NaD, despite being a
strong transition, shows absorption of only $\sim 95\%$ of continuum level and
is mostly found in small, optically thick clouds.  This suggests that the NaD is
tracing only the densest and/or dustiest regions which are more slowly
accelerated by star formation feedback than the surrounding, more dilute medium.
In fig.~\ref{fig:apert}, dusty regions in the aperture are apparent E and W of
knot C. Seen in the better resolution of fig. 1 in \citet{Adamo2010}, these
regions seem like they might be connected by a narrow dust lane partially
covering the background source. We speculate that the NaD absorption of
\citet{Sandberg2013} might be associated with this.

Also the absorption feature in the Lyman $\alpha$ profile in the upper
panel seems to morphologically follow the shape of the metal lines,
indicating that radiative transfer effects are modest, indicative of a
fairly low column density of \ion{H}{1} around line center. We find a
Lyman $\alpha$ peak velocity of
$v_{\rm peak}^{\mathrm{Ly}\alpha} = 158 \pm 1$ km s⁻¹ relative to the 
systemic velocity found by \citet{Sandberg2013}, derived from nebular emission
lines in the region around knot C. This velocity, as it is also
discussed in \citet{Verhamme2015}, is just consistent with their
theoretical predictions for a density-bounded, low-column density
system, albeit on the upper limit of their allowed range.

\subsection{Metal and \ion{H}{1} column
density}\label{metal-and-column-density}

Lyman $\alpha$ escape is mainly governed by gas at or near systemic
velocity. In the middle panel of fig.~\ref{fig:WithColDens} (a) is shown
the best-fit column density $N_{\rm Si II}$ for each velocity bin. The
value at systemic velocity is $\log_{10}(N_{\rm Si II}) = 12.1 \pm 0.2$.
We adapt the value for local starbursts of log(Si/O) $= -1.59 \pm 0.07$
from \citet{Garnett1995}, and use this to estimate the column density of
neutral Hydrogen in front of the light source in the same way as in
Puschnig et al. (submitted to MNRAS) as follows. \citet{Guseva2012}
found a metallicity of Haro 11 C of $12 + \log_{10}(\text{O/H}) = 8.1$.
With $\log_{10}$ Si/O $= -1.59 \pm 0.07$, this leads to a Si/H ratio in
the neutral medium of Si / H $= 3.24^{+0.57}_{-0.48} 10^{-6}$, leading
to a Hydrogen column density of $6.2^{+0.9}_{-1.1} \times 10^{17}$ cm⁻²
for just one velocity bin. Since \ion{H}{1} gets opaque to ionizing radiation at
$\log N \sim 17.2$ \citep{Verhamme2015}, this range is not consistent
with the low- optical depth, density bounded scenario.

Furthermore, while Ly$\alpha$ radiative transfer is dominated by gas of
$v \sim v_0$, LyC is sensitive to \ion{H}{1} at \emph{all}
velocities. The total column density of \ion{Si}{2} depends on the
configuration of the neutral clouds. Assuming that the area covered by
gas in each velocity bin is completely randomly located, a lower limit
to the total column density is:
\begin{equation}
\label{NHImean}
N_{\rm Si II}^{\rm mean} = \sum_i N_{\text{HI}, i} f_{C, i}
\end{equation}
Summing this over $-450$ km s⁻¹ $< v < 150$ km s⁻¹, the velocity range
over which the column densities can be reasonably well determined (and
removing the unphysically high value in the bin at $v \approx 222$ km
s⁻¹), yields a lower limit of $\log N_{\rm Si II} = 13.52 \pm 0.15$ and
$\log N_{\rm H I} 19.01 \pm 0.17$, corresponding to $\approx 64$ optical
depths in LyC, strongly incompatible with an optically thin,
density-bounded scenario.

However, the strong riddled ionization bounded scenario -- i.e.
consisting of only a fully neutral and a fully ionized phase -- is easily ruled
out since the Ly$\alpha$ profile does not have any appreciable emission
component at zero velocity. We therefore expect a residual neutral fraction to
remain in the ionized phase; a fraction which has a column density high enough
to block Ly$\alpha$ efficiently at line center, but low enough to be at least
part transparent to LyC radiation and undetected in \ion{Si}{2}.

We can estimate the lowest detectable $N_{\rm Si II}$ by noting that the
relative errors for the \ion{Si}{2} lines in our spectrum are
$\sim 0.05$. Assuming a covering fraction of unity for the dilute
neutral component, and adopting the oscillator strength of the strongest
of the \ion{Si}{2} transitions, $f\lambda_{1260 \AA} = 1486.8$, we find
in the limit that $I/I_0 = e^{-\tau} = 0.95$, and eq.~\ref{eq:tau}
becomes:
\begin{equation}
\label{eq:silim}
N_{\rm Si II}^{\rm min} = -\log_{e}(0.95) \frac{3.768 \times 10^{14}}{f\lambda} 
    = 10^{10.1} \text{cm}^{-2},
\end{equation}
which with the adopted metallicity for Haro 11 C corresponds to a
minimum Hydrogen column density of
$N_{\rm HI}^{\rm min} \sim 4.0 \times 10^{15}$ cm⁻². This leaves around
2 orders of magnitude in $N_{\rm H I}$, in which the gas is not detected
in \ion{Si}{2} and is optically thick to Ly$\alpha$ while translucent to
LyC. If we require the gas to be detected in at least two of
the lines included in this analysis, the limiting hydrogen column
density becomes $N_{\rm HI}^{\rm min} \sim 4.9 \times 10^{16}$ cm⁻²,
adding another order of magnitude to the allowed range, but we adopt the
lower value as a conservative estimate.

The existence of a diffuse neutral component being present in the ionized medium
seems consistent with what is found in Lyman Break Analogs and Green Pea
galaxies \citep{Heckman2011, Henry2015}.  These galaxies have sometimes very low
LIS absorption depths, which is usually indicative of low covering fractions and
high porosity of the neutral medium - and yet they find covering fractions of
near unity of \ion{H}{1} from Lyman $\beta$ absorption, indicating column
densities of $N \gtrsim 10^{16}$ cm$^{-2}$. Like in the case of Haro 11, this
indicates that a non-negligible neutral component must be present in the ionized
phase, although it may very well be of such low densities and metallicity that
metal absorption from this medium is undetectable.

The column densities we derive for \ion{Si}{2} in this work are
generally between $12.0 \lesssim \log N \lesssim 12.5$, column densities
around which the transitions involved become optically thick: For
$\lambda \lambda 1260, 1526 \text{ and } 1304$, $\tau$ becomes 1 at
$\log N \sim 11.3, 12.2$ and $12.5$, respectively. Off the regions of
strongest absorption, and in particular at $v_0$, only $\lambda 1260$
seems to be optically thick, and it seems the column densities arrived
at can be trusted. But at line center, around $v\sim v_{\mathrm{int}}$, the
column densities found are so close to the limit that they are most
likely to be interpreted as lower limits. There might also be
systematics in the determination of the continuum around the lines which
may lead to the lines at $\lambda \lambda 1304, 1526$ to be falsely seen
as shallower than $\lambda 1260$, meaning that all computed column
densities are really lower limits rather than actual values. The
confidence intervals in fig. \ref{fig:WithColDens} do \emph{not} reflect
this possible saturation, but only the formal errors from the best
approximation to the residual intensity, they do not include
systematics.

If the computed column densities are in fact lower limits, this means
the inferred \ion{H}{1} column densities are also lower limits. This
would strengthen the modified riddled, ionization-bounded medium
scenario that the ISM on the line of sight to Haro 11 C consists of
dense, neutral clumps with an ionized interclump medium containing a
dilute neutral component.

\subsection{Neutral gas metallicity}\label{neutral-gas-metallicity}

The exact value of the metallicity is however uncertain. The values
found from nebular recombination lines by \citet{Guseva2012} are
measured mainly in the central \ion{H}{2} regions around the clusters;
and differ by 0.2 dex between knot B and C. The neutral, outflowing gas
could be mixed, or have an unseen LOS distance component larger than the
knot separation, drawing into question which is the better value to
assume for this gas. We base our conclusions on the value found for knot
C, but note that using the metallicity of $12 + \log (O/H) = 8.3$ found
for knot B by \citet{Guseva2012}, the inferred \ion{H}{1} column
densities are $3.9 \pm 0.6 \times 10^{17}$ cm$^{-2}$.
However, the question of the exact metallicity of the outflowing gas is
complicated. Gas closer to the star forming regions is expected to be
more strongly enriched than gas further away, which would imply that the
\ion{H}{1} column density is \emph{larger} than inferred from
\ion{Si}{2} above. To this can be added the further complication
stemming from the merger event that the galaxy is currently undergoing,
which may have mixed gas of different metal contents. In any case,
though, we would expect the regions nearest to the starbursts to
generally have higher metallicity than the surrounding cool gas, such
that the HI column density inferred above is more likely to be
underestimated than overestimated.

\subsection{Radiative transfer effects}\label{sec:rt}

The conclusions drawn by the AOD method rest on the assumption that the
observed lines are close to being pure absorption lines, with
redistribution of photons due to radiative transfer effects being
modest. Modelling work by e.g. \citet{Prochaska2011} and
\citet{Scarlata2015} has shown that in certain conditions, radiative
transfer can re-fill absorption features in a way that can make an
isotropic, optically thin medium mimic the observational fingerprints of
a system of optically thick clumps. We therefore need to investigate
whether our conclusions could be generated by such effects.

\begin{figure}
\centering
\includegraphics[width=3.500in]{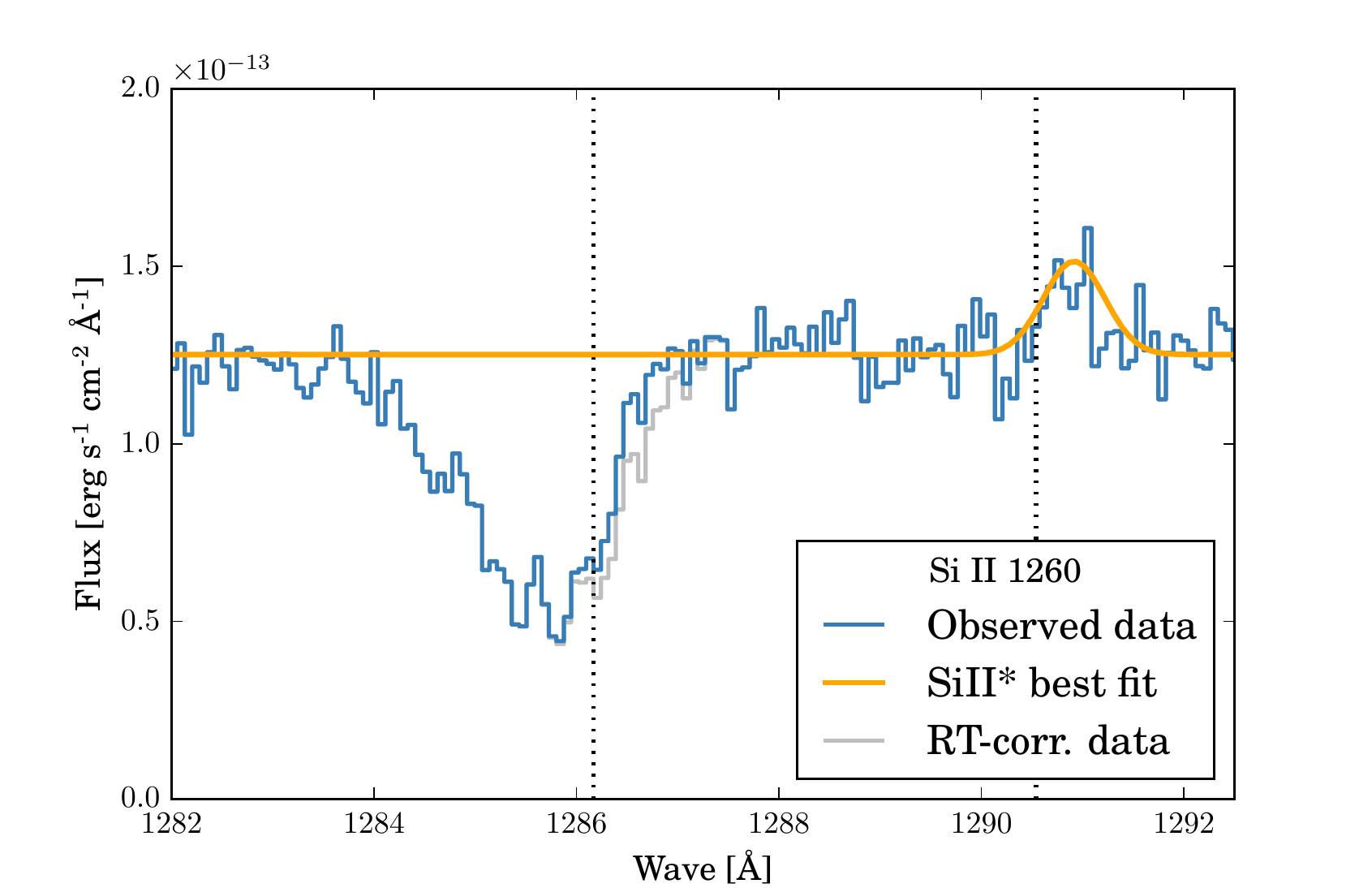}
\caption{\ion{Si}{2} $\lambda 1260$ and its fluorescent line at
$\lambda 1265$, with the observed data shown in blue. The orange curve
shows a fit to the fluorescent emission, and in gray is shown the
worst-case absorption profile corrected for radiative transfer effects
re-filling the absorption trough.}\label{fig:rt}
\end{figure}

Because the ground level in \ion{Si}{2} is a doublet with a short-lived
upper level, each of the Silicon lines included in the analysis has a
fluorescent emission companion with practically no absorption component.
The fluorescent line can be used to constrain the possible effect of
radiative infilling of the absorption trough. If each photon in the
fluorescent transition on average has scattered once in the neutral
medium, the ratio of emission in this line and re-filling of the
resonant absorption line is simply that of their Einstein coefficients
$A_{ki}$. For each subsequent scattering, more photons escape directly
through the fluorescent channel, and the infilling of the absorption
line is further suppressed.

In fig.~\ref{fig:rt}, the observed line profile is shown in blue, along
with the best fit to \ion{Si}{2}$^*$ shown in orange. Since the
fluorescent and resonant reemission originate in the same environments,
the shape of the resonant reemission profile can be completely
determined from the shape and center of the fluorescent line and the
assumption of one scattering. It is therefore simple to construct this
theoretical reemission line, subtract it from the observed absorption
line and find the limiting depth of the pure absorption line without
reemission; this modeled line is shown in gray in the figure. It is
readily seen that the difference between worst-case intrinsic and
observed line is so small that it cannot affect our conclusions about
the covering factors and column densities of the neutral gas
significantly.

We have not performed similar measurements for the lines at
$\lambda \lambda 1304, 1526$, but note that if significant refilling is
present here, this would mean the column densities had been
underestimated in our analysis, which would strengthen the conclusions
of a clumpy medium optically thick in \ion{Si}{2} $1260$, but part
transparent in the much weaker other lines.

\subsection{\ion{C}{2} $\lambda 1334$
absorption}\label{lambda-1334-absorption}

Looking back at Fig.~\ref{fig:SingleLines}, the profile of \ion{C}{2}
$\lambda 1334$ is clearly deeper than \ion{Si}{2} $\lambda 1260$, which
we had otherwise concluded is optically thick and thus provides a limit
for the velocity binned covering fractions. We believe the explanation
is a contribution from \ion{C}{2} $\lambda$ 1335.7 blending with
$\lambda 1334$. This transition has an oscillator strength of
$f_{ik} = 0.114$, compared to $f_{ik} = 0.127$ for $\lambda 1334$ (there
is a third line at 1335 Å, but that is an order of magnitude weaker), so
given a sufficient population of its ground level, it is strong enough
to make a non-negligible contribution to the resulting line profile. One
should however bear in mind that these lines arise from different lower
fine structure levels; $\lambda 1334$ arises from ²P$_{1/2}$, while the
other two arise from ²P$_{3/2}$. The population of the latter level
depends sensitively on physical conditions in its system of origin,
especially those in photo dissociation regions.

\begin{figure}
\centering
\includegraphics[width=3.500in]{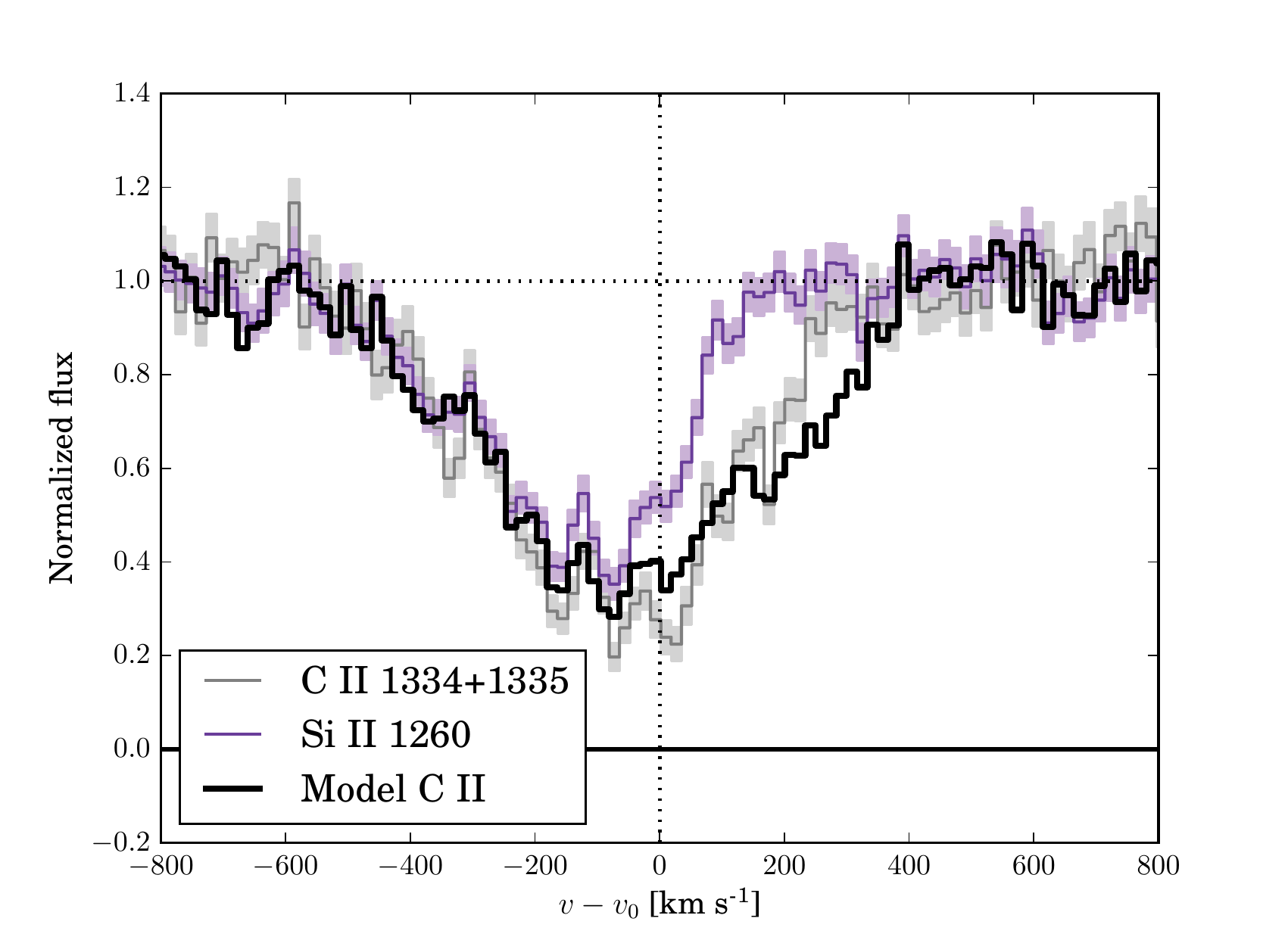}
\caption{\ion{C}{2} 1334, \ion{Si}{2} 1260, along with a \ion{C}{2}
line synthesized from \ion{Si}{2} (see text for
details).}\label{fig:CII}
\end{figure}

Fig.~\ref{fig:CII} shows a quick back-of-the-envelope check of this
hypothesis. Assuming the ions reside in the same physical regions, we
used the \ion{Si}{2} and a second contribution created by shifting the
original profile to the red by the appropriate velocity ($\sim 260$ km
s⁻¹) and multiplying it by a free parameter, which was chosen by eye to
give a reasonable replication of the observed \ion{C}{2}; the chosen
value shown in the figure is 0.7 times the original line. It is clear
from the figure that the reproduction of the observed \ion{C}{2} line is
very reasonable, considering that it is generated from a different
species with a very different ionization potential.

Whether this is a physically reasonable strength is difficult to say
with precision, but we note that to reproduce the feature observed,
$\lambda 1334$ must be optically thick and $\lambda 1335$ part
transparent. This leaves open a degeneracy of relative level populations
and ion abundance, such that the choice of this value only weakly
constrains these quantities and thus it is compatible with a wide range
of scenarios.

It is also worth noting that \ion{C}{2} has a significantly higher
ionization potential than \ion{Si}{2}, such that the former will be more
abundant in regions of higher ionization, with \ion{C}{2} in effect
tracing slightly different regions than \ion{Si}{2}. Furthermore, in
these regions of higher ionization, the relative level populations of
the Carbon ground state can be significantly altered by shifts in
e.g.~electron density and radiation field, which can alter the line
shape further. We make no attempt at mapping these complex conditions
here, but note only that there exists a range of non-exotic physical
effects which can generate the deeper and wider $\lambda 1334$ profile
we observe while still being compatible with the ISM being optically
thick in \ion{Si}{2} $\lambda 1260$.

\subsection{Conclusion}\label{conclusion}

In this work, we have re-analyzed an archival FUV HST-COS spectrum 
of Haro 11 to
investigate the kinematics and geometry of both the hot, ionized and cold,
neutral ISM along the line of sight to a strongly star forming knot suspected of
being the source of the observed LyC leakege from the galaxy. We have used the
Apparent Optical Depth method to compute column densities and velocity-binned
covering fractions of the gas and compared these results to the extreme cases of
either an optically thin, density bounded neutral medium, or a riddled,
optically thick, ionization bounded neutral medium. Assuming that the LyC
emission previously observed from this galaxy indeed does originate from knot C,
we find that the observations are not compatible with the latter case, since the
characteristic, bright Lyman-$\alpha$ emission spike at line center is absent in
this spectrum. Furthermore, the observations are not consistent with an
optically thin, density bounded neutral medium.

We confirm previous authors concluding that the medium is highly ionized
with clumps of neutral gas of low velocity-binned covering fraction,
increasing the probability of finding direct sight lines to the
background star cluster between these. The clumps have HI column
densities of gas around $v=0$ in the range
$N_{\rm HI} = 6.2^{+0.9}_{-1.1} \times 10^{17}$ cm$^{-2}$ given the 
metallicity of the background \ion{H}{2} region, which is likely a slight 
underestimate of $N_{\rm HI}$. A conservative estimate of the \ion{H}{1} column
density integrated over all velocities is
$\log N_{\rm H I}^{\rm tot} = 19.01 \pm 0.17$ cm$^{-2}$. There is a possibility
that the found column densities are in fact lower limits, since the
found \ion{Si}{2} column densities are close to the limit where this ion
becomes optically thick. We therefore conclude that the leaked ionizing
photons, if originating from this cluster, most likely escaped via sight
lines between the neutral clumps, through an ionized medium which must
contain a neutral gas column density
$5 \times 10^{13} \lesssim N_{\rm HI}^{\rm HIS} \lesssim 4 \times 10^{15}$
cm$^{-2}$. This range is bounded downwards by the value at which HI becomes
transparent to Lyman-$\alpha$, and upwards by the detectability of
\ion{Si}{2} in this observation.

It is interesting to note that Haro 11 is relatively on neutral 
gas poor, and has a very strong ongoing starburst episode. If we were to see a 
case in which neutral gas depletion and/or ionization provided a 
density-bounded scenario, this would be a likely place to find it. 
Still, it seems that dynamical effects of feedback and galaxy-scale 
interaction plays the greater part in peeling away the neutral gas from the 
central starbursts and opening passages for Ly$\alpha$ and LyC to 
escape. Given that star formation and merger activity is significantly stronger 
at $z\gtrsim 1$ than in the local Universe, and that Dark Matter potential wells
were shallower, these mechanisms could be even more important in that era. 
one could speculate that this could provide a piece of the puzzle of which
sources have driven the epoch of reionization.


\section*{Acknowledgements}\label{acknowledgements}
\addcontentsline{toc}{section}{Acknowledgements}

The authors thank the anonymous referee for constructive and insightful 
comments, which have helped improve the quality of this paper significantly. 

GÖ \& MH acknowledge the support of the Swedish Research Council,
Vetenskapsrådet, and the Swedish National Space Board (SNSB). MH is an
Academy Fellow of the Knut and Alice Wallenberg Foundation. This project
has made extensive use of the Python-based packages Numpy \citep{Numpy},
SciPy \citep{SciPy}, Matplotlib \citep{Matplotlib}, Pandas
\citep{Pandas}, LMfit \citep{lmfit2014}, and Astropy
\citep{Astropy2013}.

\bibliography{./main.bib}

\end{document}